\documentclass[preprint,12pt]{elsarticle}

\usepackage{amssymb}
\usepackage{amsmath}

\usepackage{lineno}

\journal{Journal of Subatomic Particles and Cosmology}

\usepackage{xspace}

\newcommand{\tttt}{\ensuremath{\text{t} \bar{\text{t}} \text{t} \bar{\text{t}}}\xspace}
\newcommand{\ttH}{\ensuremath{\text{t}\bar{\text{t}}\text{H}}\xspace}
\newcommand{\tttW}{tttW\xspace}
\newcommand{\tttq}{\ensuremath{\text{ttt}q}\xspace}

\usepackage{caption}
\biboptions{sort&compress}

\usepackage{hyperref}
\usepackage{url}

\begin{document}

\begin{frontmatter}



\title{Search for Physics beyond the Standard Model in four- and three-top quark production}


\author{Maryam Shooshtari on behalf of the CMS Colaboration} 

\affiliation{organization={Austrian Academy of Sciences},
            city={Vienna},
            country={Austria}}
\ead{maryam.shooshtari@cern.ch}

\fntext[copy]{Copyright 2026 CERN for the benefit of the CMS Collaboration. Reproduction of this article or parts of it is allowed as specified in the CC-BY-4.0 license.}

\begin{abstract}
A reinterpretation of four-top quark (\tttt) production is presented using the full Run~2 dataset recorded by the CMS experiment, corresponding to an integrated luminosity of 138\,fb$^{-1}$. The analysis targets BSM scenarios using the existing \tttt production measurement, including constraints on effective field theory (EFT) operators, top-philic heavy resonances, and the top-Yukawa coupling. The results provide competitive limits on several new physics models and demonstrate the sensitivity of multi-top final states to a wide range of BSM effects.
\end{abstract}




\end{frontmatter}



\section{Introduction}

Four-top production in Standard Model (SM) has been observed by ATLAS~\cite{ATLAS:Detector-2008} and CMS~\cite{CMS:Detector-2008} experiments, with CMS measuring a cross-section about 34\% higher than the SM prediction~\cite{CMS:TOP-22-013,vanBeekveld:2022hty}, motivating Beyond Standard Model (BSM) interpretations. Both ATLAS and CMS also observed high anti-correlation in production cross-section of four-top and three-top, effectively making it impossible to distinguish whether an excess is caused by one or the other. These proceedings summarize the CMS search for new physics in \tttt and ttt production using the 2016–2018 dataset of 138~fb$^{-1}$ at $\sqrt{s}=13$~TeV, interpreted in three complementary frameworks:
Standard Model Effective Field Theory (SMEFT), searches for simplified top-philic heavy resonances, and study of the structure of the top-quark Yukawa coupling~\cite{CMS-PAS-TOP-24-008}.

\section{Analysis setup}

This analysis targets final states with two same-sign, three, or four leptons. These are divided into exclusive signal regions (SRs) and control regions (CRs) based on jet and b jet multiplicity, the scalar sum of all jet $p_{\mathrm{T}}$ in the event ($H_{\mathrm{T}}$), and the presence of a Z boson candidate. A boosted decision tree (BDT) classifier has been trained on the dilepton, and separately three and four lepton SRs with the highest jet multiplicity to separate \tttt-like events from the background~\cite{CMS:TOP-22-013}. 

\begin{figure}[htp]
\centering
\includegraphics[width=0.45\textwidth]{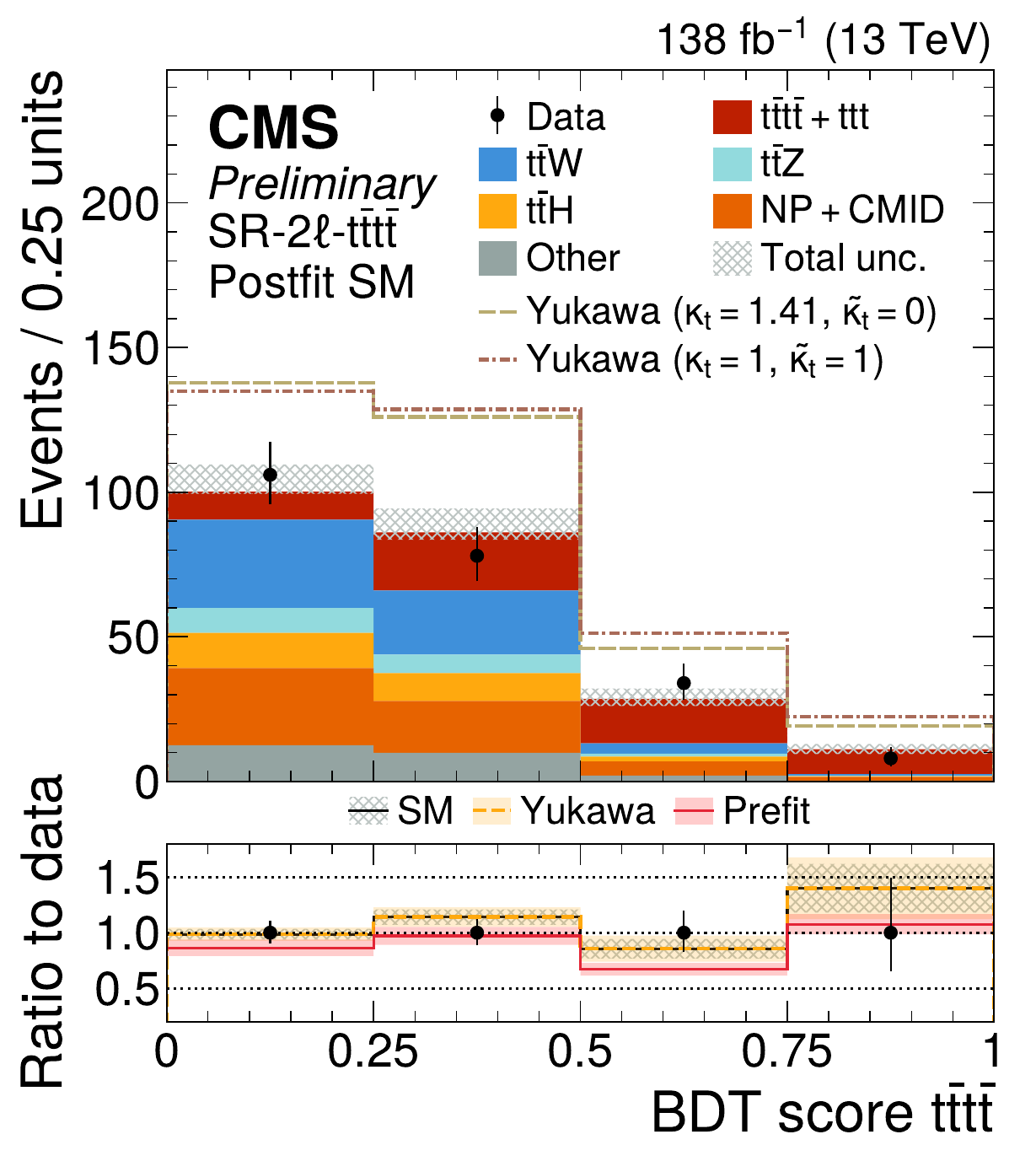}%
\hspace*{0.02\textwidth}
\includegraphics[width=0.45\textwidth]{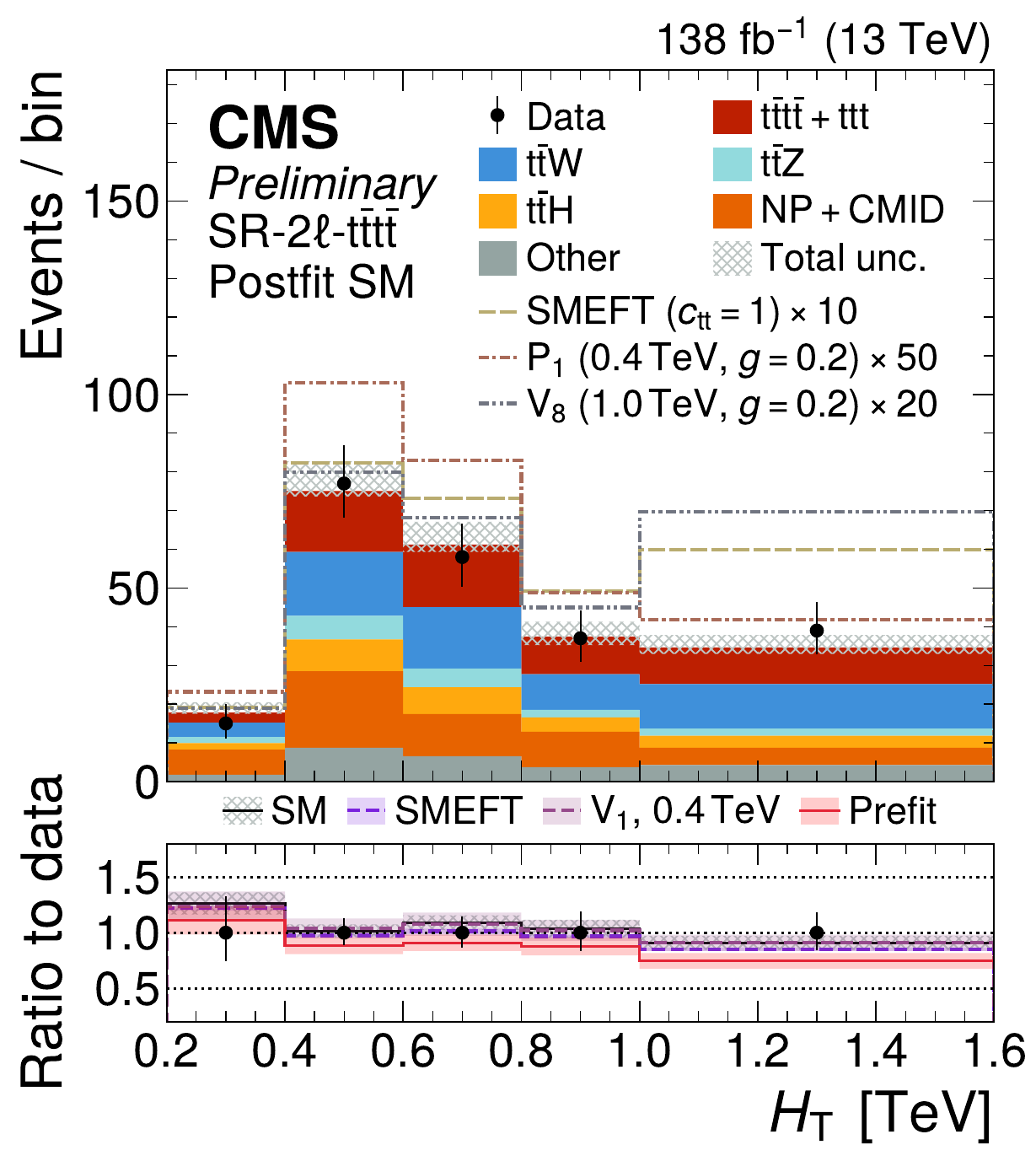} 
\caption{Postfit BDT score \tttt (left) and $H_{\mathrm{T}}$ (right) distributions in the dilepton SR, with the ratio panels for SM and BSM hypotheses~\cite{CMS-PAS-TOP-24-008}.
}
    \label{fig:postfit}
\end{figure}

Figure~\ref{fig:postfit} shows two possible input distributions. For the SM and the Yukawa fit, the BDT score was used, as it offers the best signal-background separation. For the SMEFT and heavy resonances fit, $H_{\mathrm{T}}$ distribution is used as to exploit the highest tails of the energy where the effects from these scenarios are enhanced.

\section{Standard Model Effective Field Theory}
The SMEFT framework extends the Lagrangian of the SM with new terms, including dimension 6 operators that describe new physics effects~\cite{Degrande:2012wf,Aoude:2022deh,Zhang:2017mls,Aleshko:2023rkv,Aleshko:2025jua,DiNoi:2025uhu}. The analysis considered four different four-fermion operators ($\mathcal{O}_{tt}$, $\mathcal{O}^{1}_{QQ}$, $\mathcal{O}^{1}_{Qt}$, $\mathcal{O}^{8}_{Qt}$) and a complex operator modifying top-Higgs interactions ($\mathcal{O}_{t\phi}$). Independent Monte Carlo (MC) simulations of EFT effects in signal processes \tttt , ttt, and \ttH were produced at leading-order (LO) using MadGraph5\_aMC@NLO~\cite{Alwall:2014hca} (MG5$\_$aMC) and the top quark decays simulated with MADSPIN~\cite{Artoisenet:2012st} using SMEFTsim~3.0 convention~\cite{Brivio:2017btx, Brivio:2020onw}. 

\begin{figure}[t!]
\vspace{-0.4cm}
\centering
\includegraphics[width=0.45\textwidth]{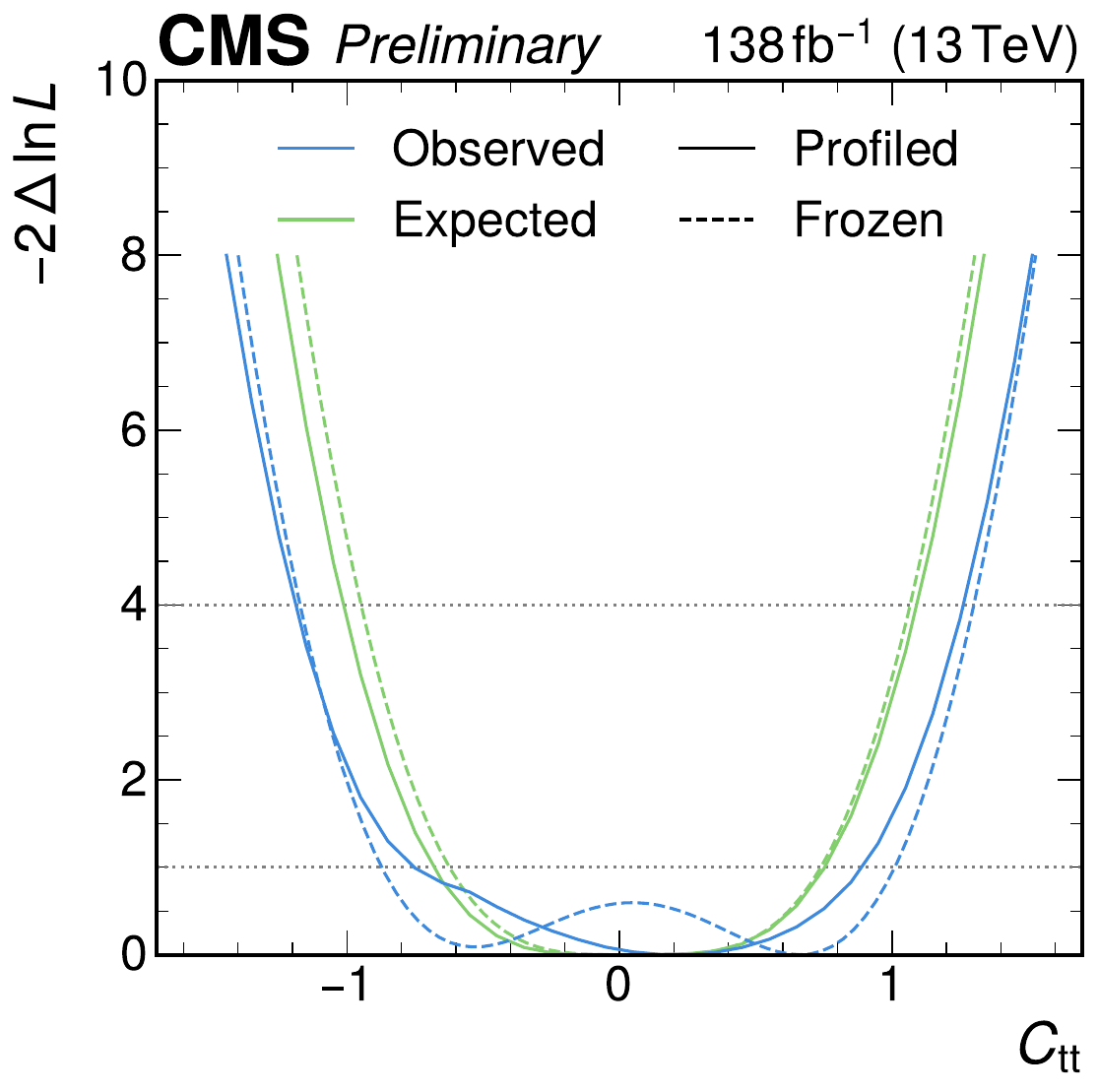}
\hspace*{0.05\textwidth}%
\includegraphics[width=0.45\textwidth]{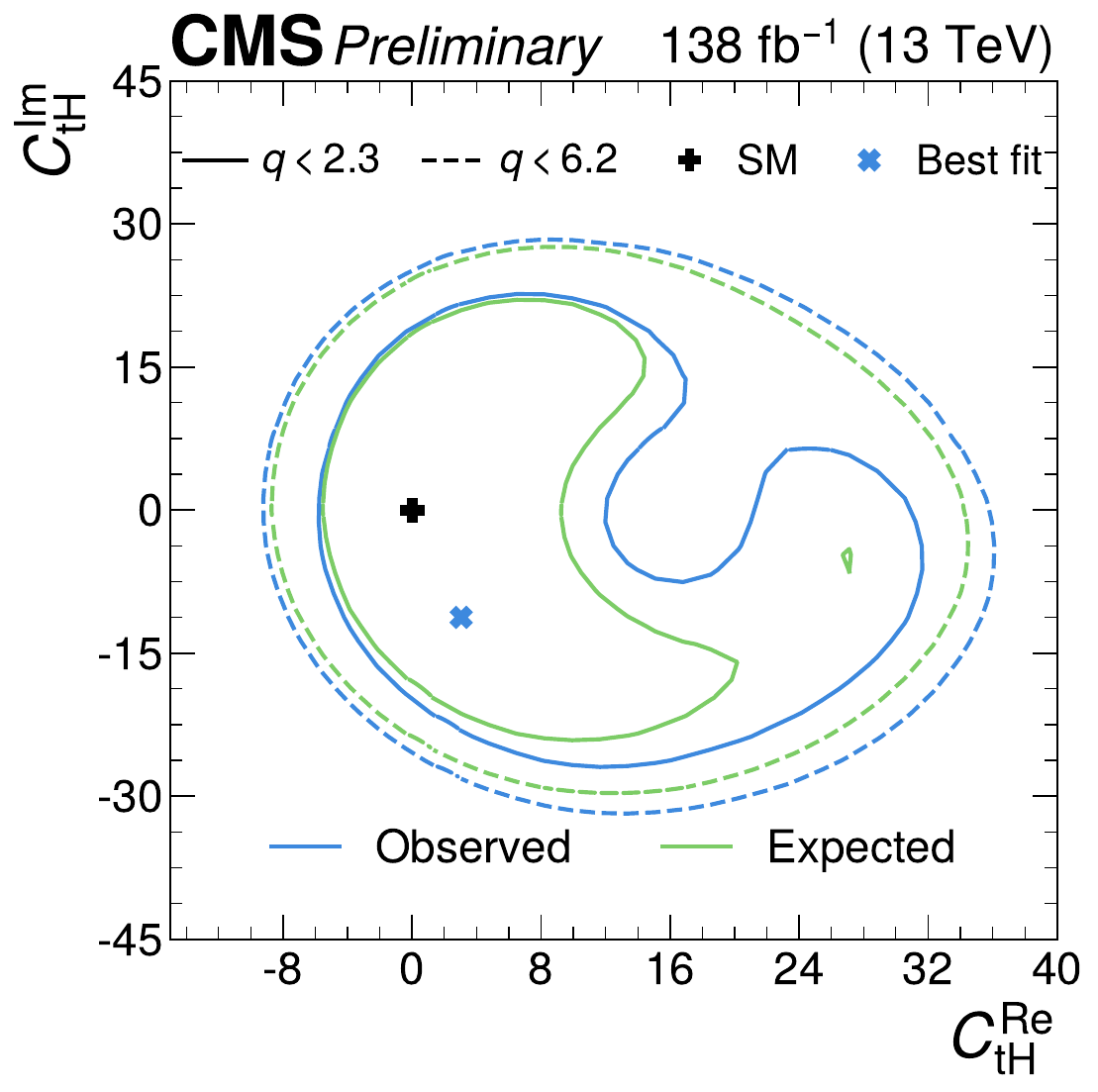} 
\caption{Negative log-likelihood difference from the best fit value for the one-dimensional scan of $c_{tt}$ (left), and the two-dimensional scan of the $c_{tH}^{Re}$ and $c_{tH}^{Im}$ (right) showing the structure of the Yukawa coupling~\cite{CMS-PAS-TOP-24-008}.}
\label{fig:eft}
\end{figure}

A six dimensional fit, considering 4 four-fermion Wilson Coefficients and the real and imaginary parts of the $\mathcal{O}_{t\phi}$ as parameters of interest, was performed to obtain best values and the corresponding 68\% and 95\% confidence intervals for each of them. Figure~\ref{fig:eft} shows some results obtained. 

\section{Top-philic heavy resonances}
\begin{figure}[t!]
\vspace{-0.4cm}
\centering
\includegraphics[width=0.45\textwidth]{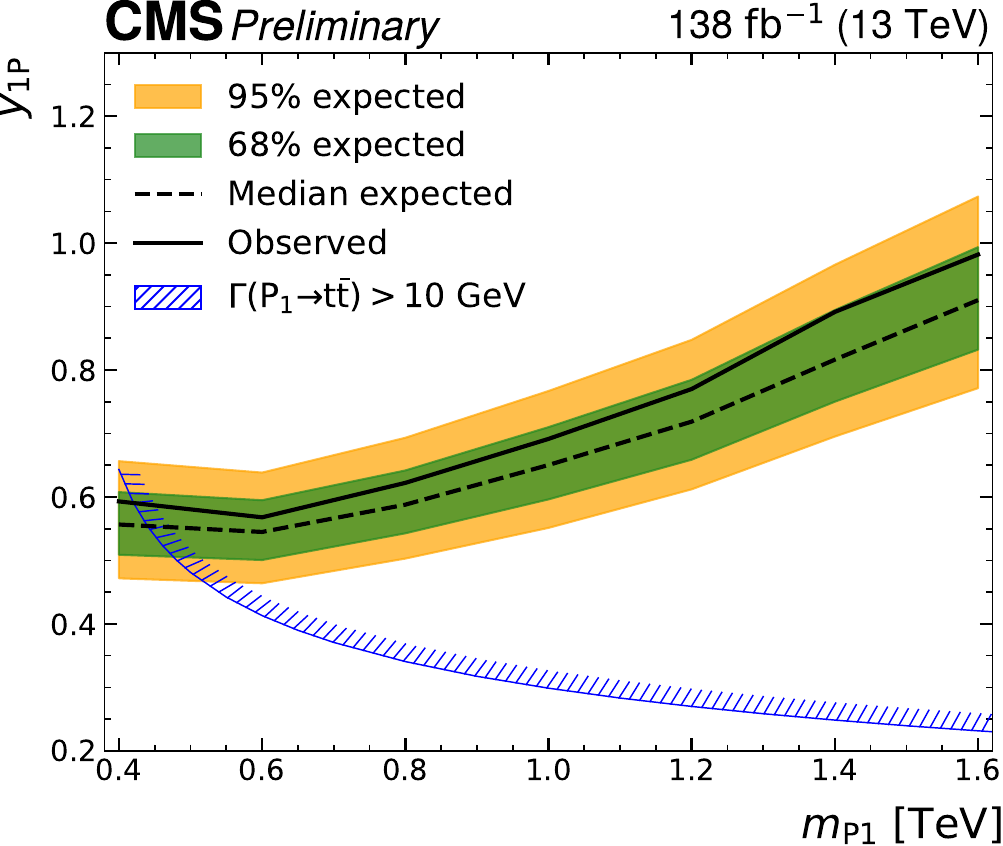} 
\hspace*{0.02\textwidth}%
\includegraphics[width=0.45\textwidth]{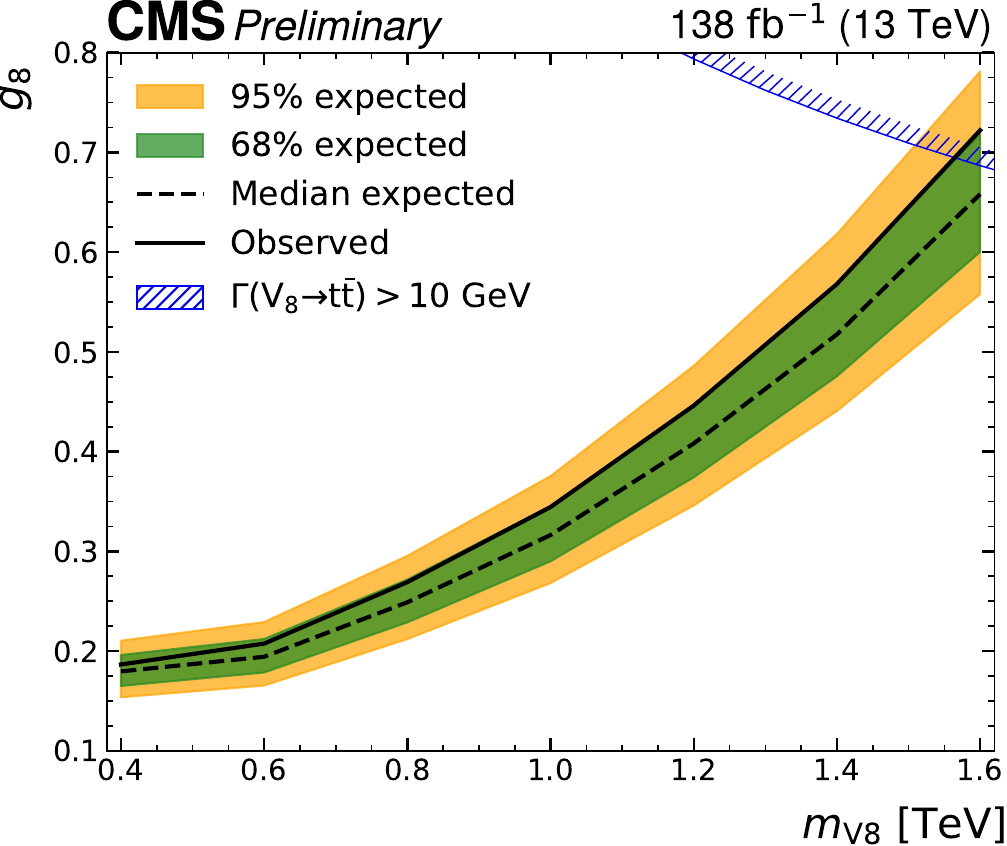} 
\caption{Upper limits found for coupling of a pseudoscalar singlet (left) and a vector octet (right) boson as a function of their mass~\cite{CMS-PAS-TOP-24-008}. The blue line shows the limit where the partial decay width of the boson would be higher than 10~GeV and our simulation is no longer valid.}
\label{fig:bosons}
\end{figure}

The second section of the analysis explores production of \tttt or ttt events through heavy top-philic bosons. Six  scalar, pseudoscalar, and vector bosons in color singlet or octet state models are considered~\cite{Darme:2021gtt}. MC simulations for these were produced at LO using MG5$\_$aMC for seven different boson masses in the range of 400~GeV to 1.6~TeV, with a fixed decay width of 10~GeV. Using these, we obtained upper limits of the coupling strength as a function of boson mass. Figure~\ref{fig:bosons} the results for the pseudoscalar singlet (left), and vector octet (right) model.  

\section{Top-Yukawa coupling}
\begin{figure}[h!]
\centering
\includegraphics[width=0.45\textwidth]{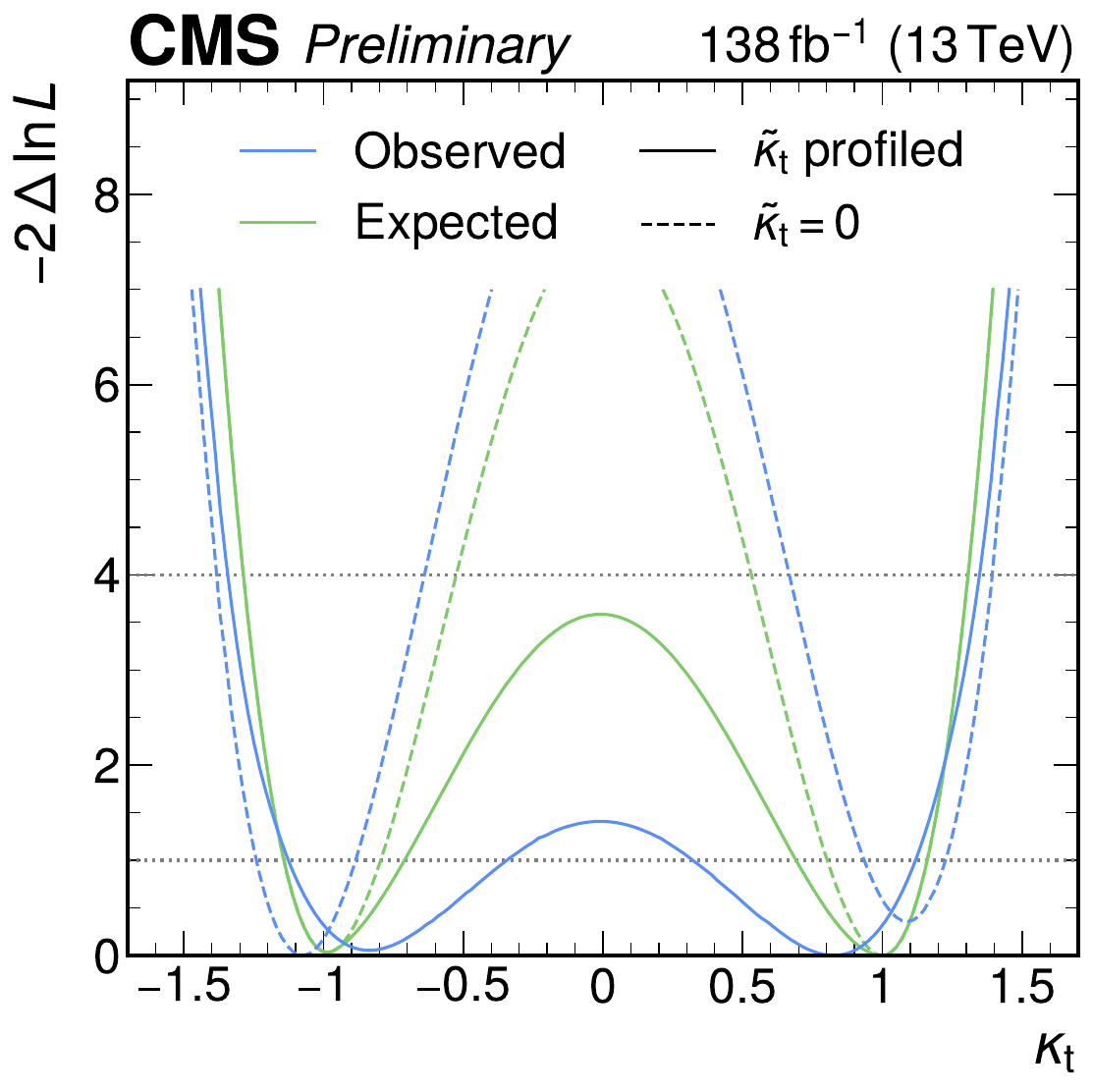}
\hspace*{0.05\textwidth}%
\includegraphics[width=0.45\textwidth]{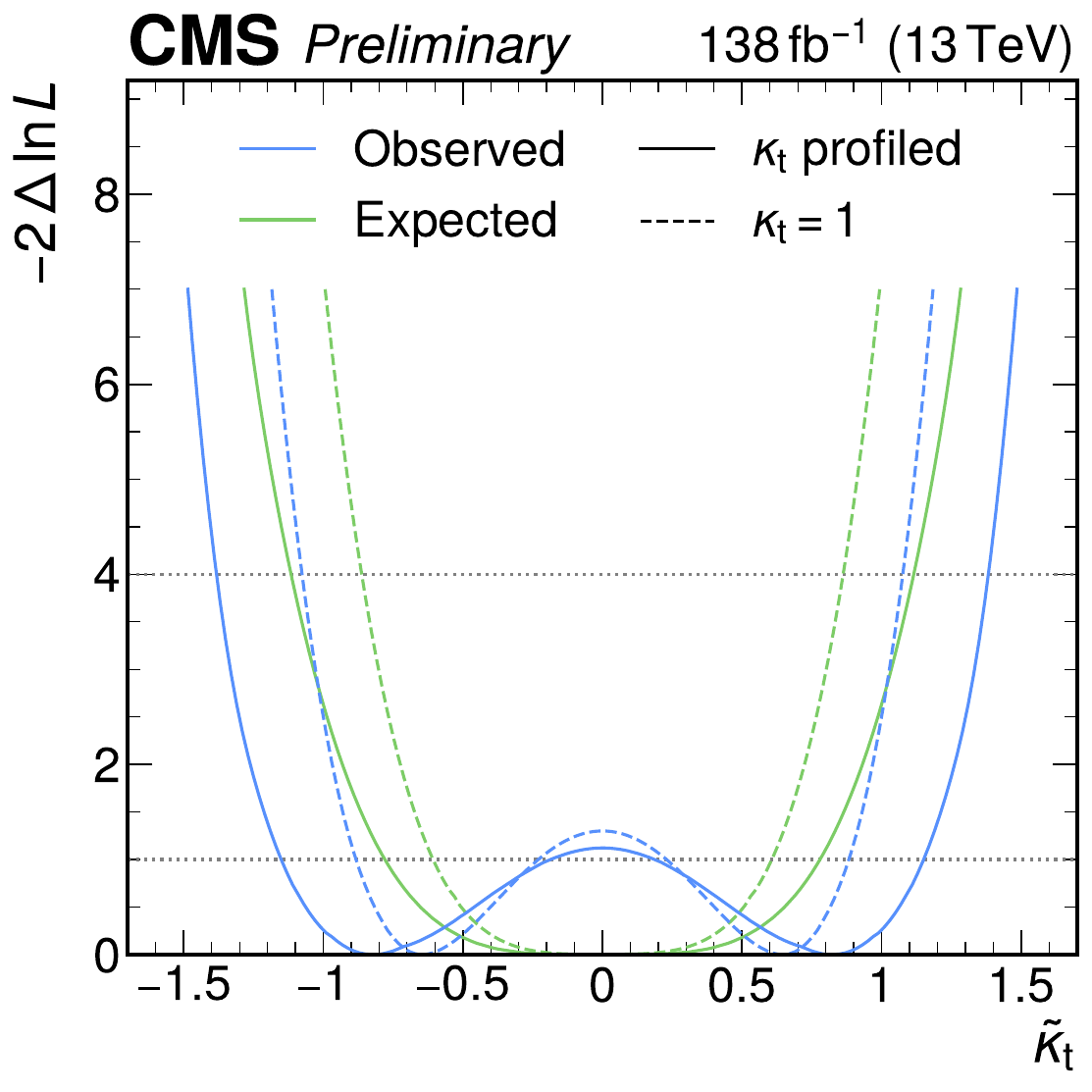} \\
\caption{One-dimensional likelihood scans of the even (left) and odd (right) Yukawa coupling modifiers~\cite{CMS-PAS-TOP-24-008}. }
\label{fig:Yukawa}
\end{figure}

High-energy multi-top events provide great sensitivity to top--Higgs interactions. Since four-top production is independent of Higgs width, it provides a great probe of the top Yukawa coupling. In \tttt, ttt, and \ttH production, electroweak contributions involving the top--Higgs interaction depend on the CP-even ($\kappa_t$) and CP-odd ($\tilde{\kappa}_t$) modifiers~\cite{Gritsan:2016hjl}. We use models from Ref.~\cite{ Demartin:2014fia} with MG5$\_$aMC to simulate LO samples with this parametrization and record the cross-section modification. We observe that \tttt production is symmetric in $\kappa_t$, while \tttW and \tttq show a sign dependence.
The results in Fig.~\ref{fig:Yukawa} express the structure of the top-Yukawa coupling. 

\section{Conclusions}
A comprehensive analysis of multi-top-quark production using 138~fb$^{-1}$ of CMS data was presented. No significant deviation from the SM expectations is observed, and the results yield improved limits on multiple classes of new-physics effects involving top quarks.

\bibliographystyle{unsrt}
\let\oldbibitem\bibitem
\renewcommand{\bibitem}[1]{\oldbibitem{#1}\vspace{-1mm}}



\end{document}